# Development of digital sideband separating down-conversion for Yuan-Tseh Lee Array

Chao-Te Li*, Derek Kubo, Jen-Chieh Cheng, John Kuroda, Ranjani Srinivasan, Solomon Ho, Kim Guzzino, Ming-Tang Chen
Institute of Astronomy and Astrophysics, Academia Sinica, Taipei, Taiwan

## ABSTRACT

This report presents a down-conversion method involving digital sideband separation for the Yuan-Tseh Lee Array (YTLA) to double the processing bandwidth. The receiver consists of a MMIC HEMT LNA front end operating at a wavelength of 3 mm, and sub-harmonic mixers that output signals at intermediate frequencies (IFs) of 2–18 GHz. The sideband separation scheme involves an analog 90° hybrid followed by two mixers that provide down-conversion of the IF signal to a pair of in-phase (I) and quadrature (Q) signals in baseband. The I and Q baseband signals are digitized using 5 Giga sample per second (Gsps) analog-to-digital converters (ADCs). A second hybrid is digitally implemented using field-programmable gate arrays (FPGAs) to produce two sidebands, each with a bandwidth of 1.6 GHz. The 2 × 1.6 GHz band can be tuned to cover any 3.2 GHz window within the aforementioned IF range of the array. Sideband rejection ratios (SRRs) above 20 dB can be obtained across the 3.2 GHz bandwidth by equalizing the power and delay between the I and Q baseband signals. Furthermore, SRRs above 30 dB can be achieved when calibration is applied.

## 1. INTRODUCTION

The Yuan-Tseh Lee Array (YTLA), formerly known as the Array for Microwave Background Anisotropy (AMiBA) [1], is an interferometric array that was originally designed to detect finite differences in the cosmic microwave background (CMB). The receiver, operating in the W-band from 86 to 102 GHz, where foreground contamination is minimal, is intrinsically single-sideband with a high-pass filter located in front of the subharmonic mixers. Because the frequency spectrum of the CMB resembles that of thermal radiation, the YTLA was equipped with a wideband analog correlator with a coarse frequency resolution to process the entire intermediate frequency (IF) band between 2 and 18 GHz to achieve high sensitivity [2]. As observations of Sunyaev–Zeldovich clusters taper off, a new initiative was launched for the YTLA to map the CO intensity for tracing the large-scale structure of star-forming galaxies in the early universe [3]. Intensity mapping images the aggregated emission from low-luminosity galaxies on very large scales without requiring the detection of individual galaxies. Two-dimensional intensity mapping has been successfully used for measuring the CMB. However, the broad spectral bands integrate the emission over redshift. The array is being retrofitted with a digital correlator to obtain spectral measurements incorporating redshift information. To probe more space volume, a down-conversion method involving digital sideband separation was adopted to double the processing bandwidth.

For heterodyne receivers used in radio astronomy, dual-sideband (2SB) architecture is preferred. In this architecture, the two sidebands are output on different ports to avoid spectral confusion and to suppress noise from the other sideband. The typical configuration of a 2SB or image rejection mixer [4], shown in Fig. 1, consists of a radio frequency (RF) hybrid, mixers used as down-converters, and an IF hybrid. The RF hybrid splits the RF signal into the in-phase (I) and quadrature (Q) signals with a 90° phase difference, and these two signals are down-converted by the mixers, which are driven by the same local oscillator (LO) signals obtained from a power splitter. The down-converted signals are then recombined by the IF hybrid to produce the lower sideband (LSB) and upper sideband (USB). However, broadband sideband-separating down-converters are difficult to build because they require the construction of two hybrids and two parallel mixer/amplifier chains with excellent amplitude and phase balance over the entire bandwidth. The gain and phase imbalances of analog components in the two parallel signal paths limit the sideband rejection ratio (SRR) to approximately 10–20 dB [5].

*ctli@asiaa.sinica.edu.tw

An increase in the speed of digital hardware has made it feasible to implement an IF hybrid by using digital signal processing (DSP). Murk et al. [6] reported a receiver with digital sideband separation. The IF hybrid was digitally implemented by using a field-programmable gate array (FPGA) to process a bandwidth of 2 × 500 MHz in real time with an image rejection between 10 and 25 dB. Morgan et al. [7] constructed a calibrated digital sideband-separating mixer (DSSM) and corrected for imbalances through offline digital processing. They processed 2 × 250 MHz of bandwidth, down-converted from the L-band (1.2–1.7 GHz), with an SRR higher than 50 dB. Finger et al. [8] implemented a real-time calibrated digital sideband-separating spectrometer to process an IF bandwidth of 500 MHz per sideband with a 4 GHz analog front end. Rodriguez et al. [5] applied the technique to build a millimeter-band sideband-separating receiver with an IF bandwidth of 2 × 500 MHz. In this report, we present a digital sideband separating down-conversion method using 5 Giga sample per second (Gsps) analog-to-digital converters (ADCs). The goal is to process a bandwidth of 2 × 2 GHz. The current design is running with a bandwidth of 1.6 GHz per sideband, constrained by timing within the FPGA.

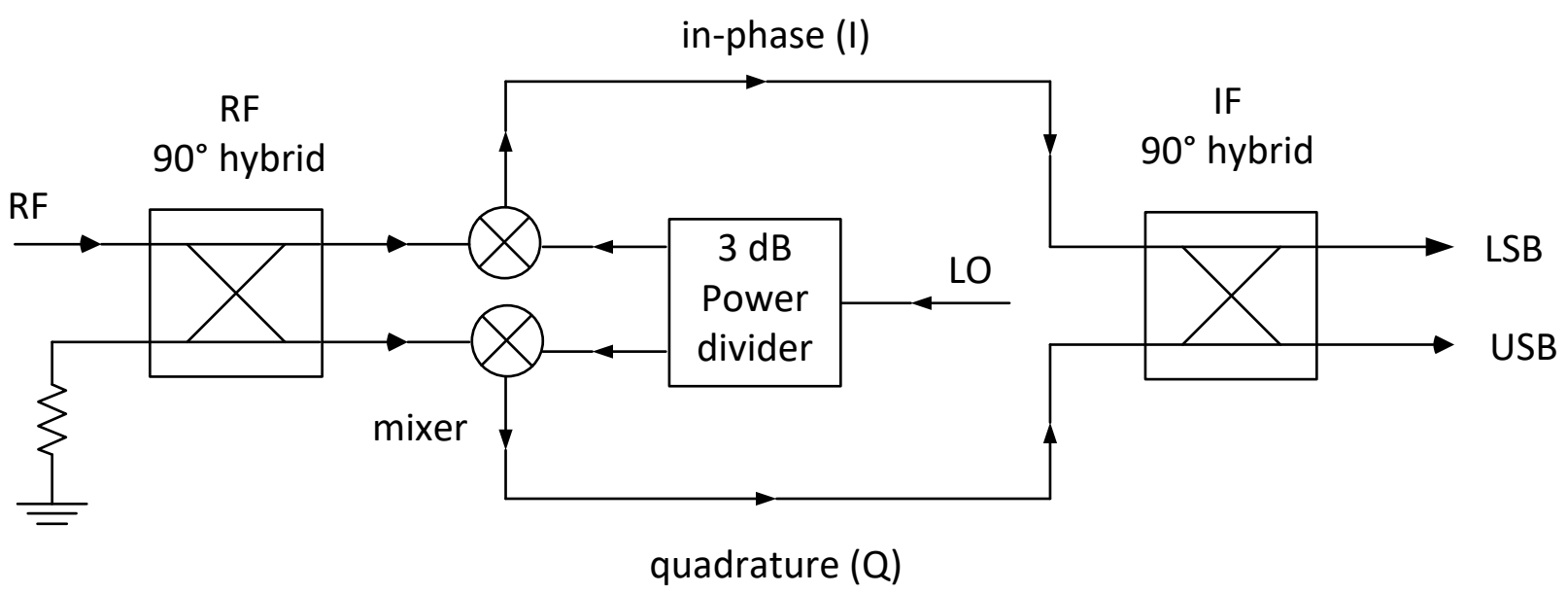

Figure 1. Typical configuration of a sideband-separating or dual-sideband down-converter.

## 2. DIGITAL SIDEBAND SEPARATION SETUP

Alternatively, for 2SB down-conversion, similar to the DSSM, the RF input can be divided in phase while the LO is divided in quadrature, as shown in Fig. 2. For this configuration, the LO phase shift $\varphi_{LO}$ should be constant with IF. However, in practice, the phase shift may show small variations because of imperfections on the RF splitter and the mixers. For calibrated digital sideband separation, the mixer outputs are digitized using ADCs, and the time domain sequences are Fourier transformed to frequency spectra. The outcomes of the frequency channels are multiplied by complex coefficients $C_1$ to $C_4$ and recombined to obtain two sidebands. When the coefficients are selected carefully, the behavior of an ideal IF hybrid can be reproduced and furthermore the phase and amplitude imbalances of both signal paths can be calibrated.

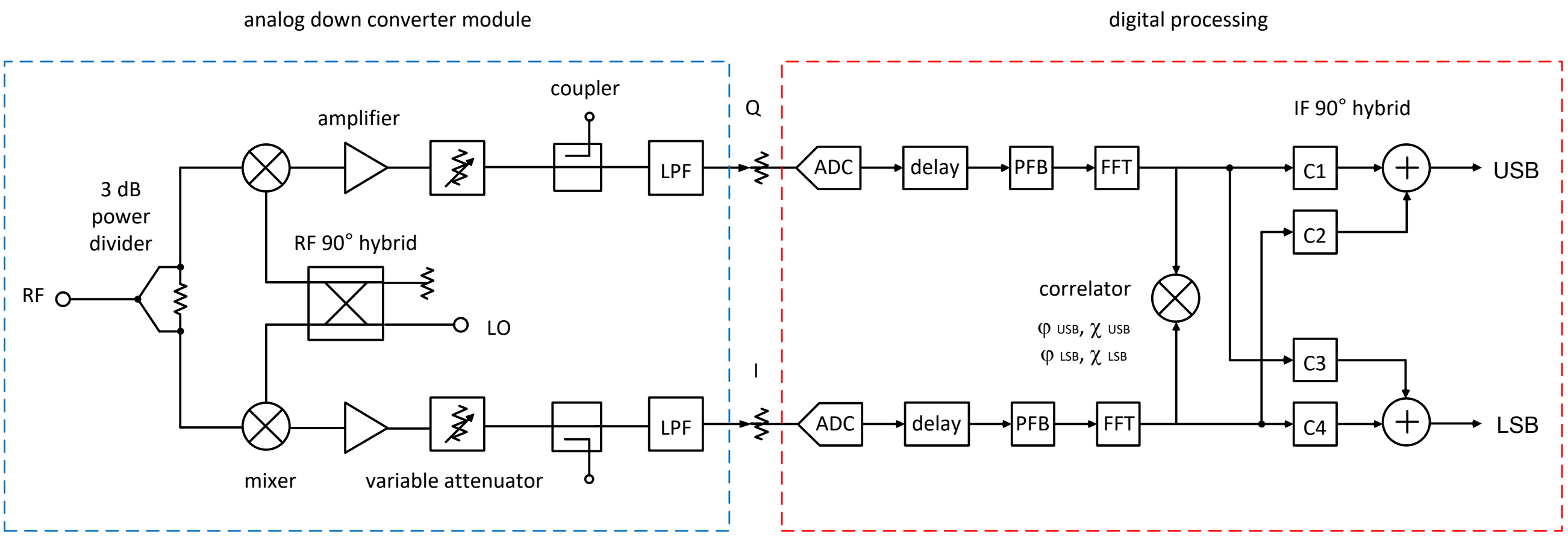

Figure 2. Block diagram of a digital sideband-separating mixer.

## 2.1 Analog down-converter module

A schematic of an analog down-converter module is shown on the left side of Fig. 2, and a photograph of the prototype is shown in Fig. 3. A broadband 90° hybrid coupler with an amplitude imbalance of ±0.5 dB and a phase imbalance of ±10° from 1 to 18 GHz was used to provide in-phase and quadrature LO signals to the mixers. The LO frequency can be tuned to position the 2 × 1.6 GHz band anywhere within the 2–18 GHz IF band of the array. A 1.5–18 GHz power splitter was used to split the signal from the receiver. Broadband balanced mixers with an RF/LO range of 2–18 GHz and an IF up to 6 GHz were used to down-convert the receiver signal to in-phase (I) and quadrature (Q) ones. The following amplifiers, with a typical gain of 30 dB from 2.5 MHz to 2.5 GHz, and the variable attenuators were used to optimize and balance the power of the I and Q signals to the ADCs. Furthermore, 10-dB directional couplers were used for power monitoring and adjustment of the variable attenuators, and low-pass filters were used to avoid signals aliasing at the ADCs.

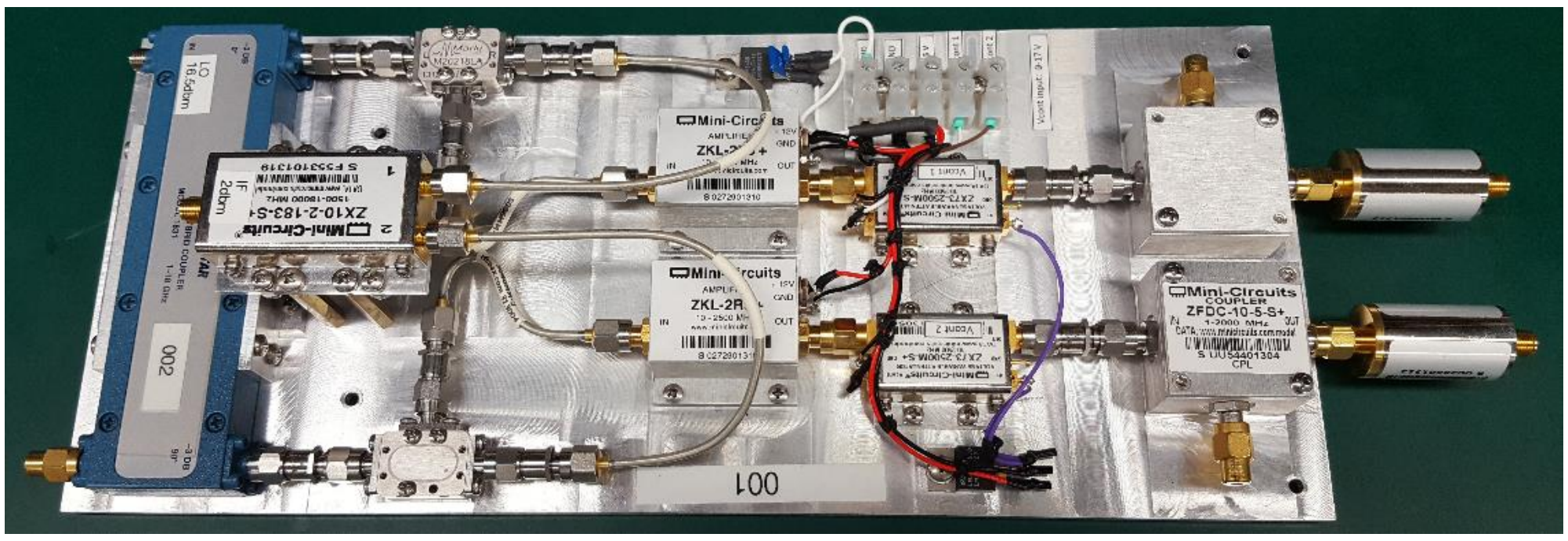

Figure 3. Photo of the prototype analog down-converter module. From left to right - the LO quadrature hybrid, the RF two-way power splitter, mixers, amplifiers, variable attenuators, directional couplers, and anti-aliasing filters.

## 2.2 Digital hybrid

The 2nd hybrid was digitally implemented using ADC boards and DSP modules, as shown in Fig. 4. Each ADC board was equipped with a 5 Gsps ADC [9] that consisted of four 8-bit ADC cores, clocked by an external clock. Each ADC core had a maximum sampling rate of 1.25 Gsps. To maximize the bandwidth, the quad ADC was operated in the one-channel mode, in which all four ADC cores are interleaved. The DSP platform used was the Reconfigurable Open Architecture Computing Hardware v.2 (ROACH2), developed by the Collaboration for Astronomy Signal Processing and Electronics Research (CASPER) [10]. ROACH2 is a stand-alone computing platform equipped with a Xilinx Virtex-6 FPGA. Z-Dok connectors were used to connect the ADC modules to the ROACH2 for data acquisition.

Time sequences from the ADCs were converted to the frequency domain through polyphase filter bank (PFB) and fast Fourier transform (FFT) in the ROACH2. There were 1024 frequency channels after the PFB/FFT. The PFB was adopted to prevent scalloping loss and leakage resulting from FFT [11]. A PFB can produce a flat frequency channel response and better suppression of out-of-band signals. To build the PFB/FFT, the CASPER library IP blocks were used. A schematic of the digital hybrid design is shown on the right side of Fig. 2. The ADC interface in the FPGA utilizes Xilinx ISERDES [12], a deserializer or serial-to-parallel converter for de-multiplexing (demux) data streams from the ADC. Xilinx ISERDES facilitates high-speed data transfer without requiring the FPGA fabric to match the input data rate. When operated in the one-channel mode, the ADC provides a demux by 4 and each of the 4 data streams are further demuxed by 4 by ISERDES in the FPGA. To achieve a 2 GHz bandwidth per sideband, the FPGA system clock would operate at 250 MHz, which poses a stringent timing constraint for large designs. Therefore, additional demux by 2 could be implemented, and the system clock rate would assume a modest value of 125 MHz.

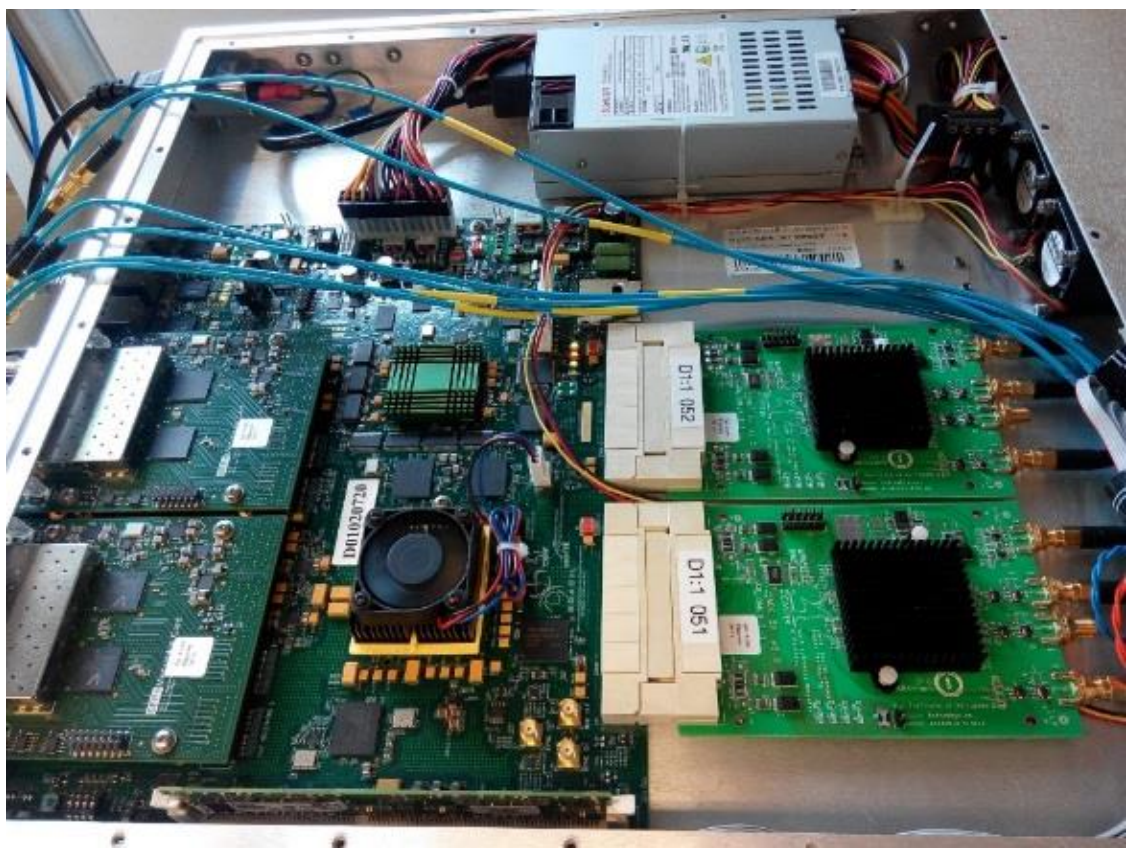

Figure 4. Digital hybrid implemented using the ROACH2 DSP module on the left and two 5 Gsps ADC boards on the right.

Following the PFB/FFT, the I and Q signals were correlated in the FPGA by using a pocket correlator [13]. The delay blocks are used to synchronize the I and Q channels, i.e. to adjust the relative delay between two channels to remove any phase wrap seen from the cross-correlations, as shown in Fig. 5. The correlation results could also be used to derive coefficients to compensate for phase and amplitude imbalance to improve the SRRs, as described in the following section.

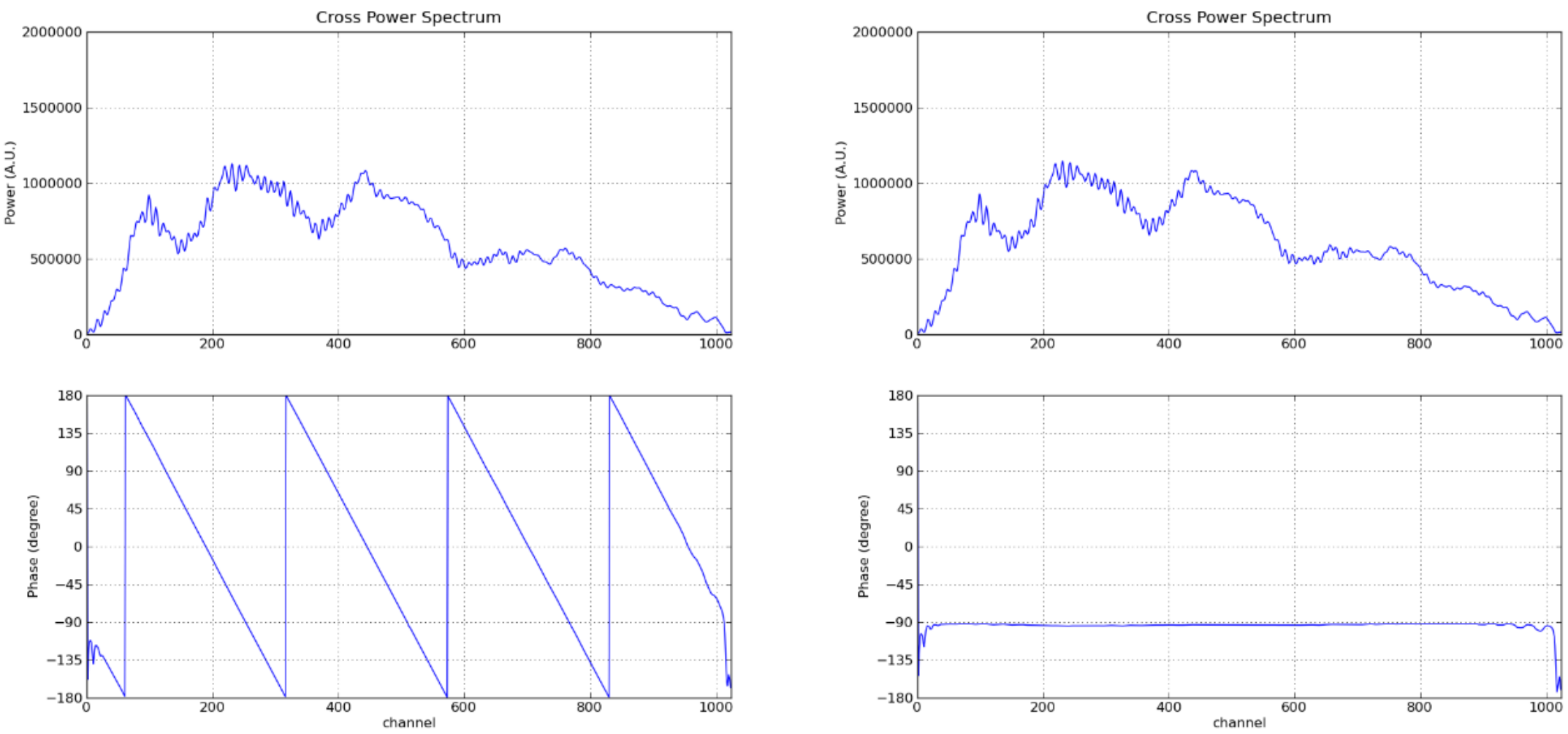

Figure 5. Band-limited noise was input into the analog down-converter module and-cross-correlations were measured between the I and Q signals. The variable delays in the FPGA were adjusted to remove phase wraps. The plots represent the amplitude and phase of the cross-correlation before (left) and after (right) the delay adjustment.

## 3. CALIBRAION AND LABORATORY TESTS

With a band-limited noise in the USB input into the analog down-converter module, the power and delay between the I and Q channels were equalized. Using nominal coefficients of the digital hybrid, namely, $C_1 = C_4 = 1 + 0\,j$ and $C_2 = C_3 = 0 - 1\,j$, the desired signal appeared at the USB output as shown in Fig. 6. To measure the SRR, the system was tested with a continuous wave (CW) signal for the 1024 channels of each sideband, as shown in Fig. 7. The SRR as defined here is, for a CW signal in a given sideband, the power ratio of the signal delivered to the desired output to that delivered to the un-desired one. The un-calibrated SRRs were measured across the two sidebands, as shown in Fig. 8.

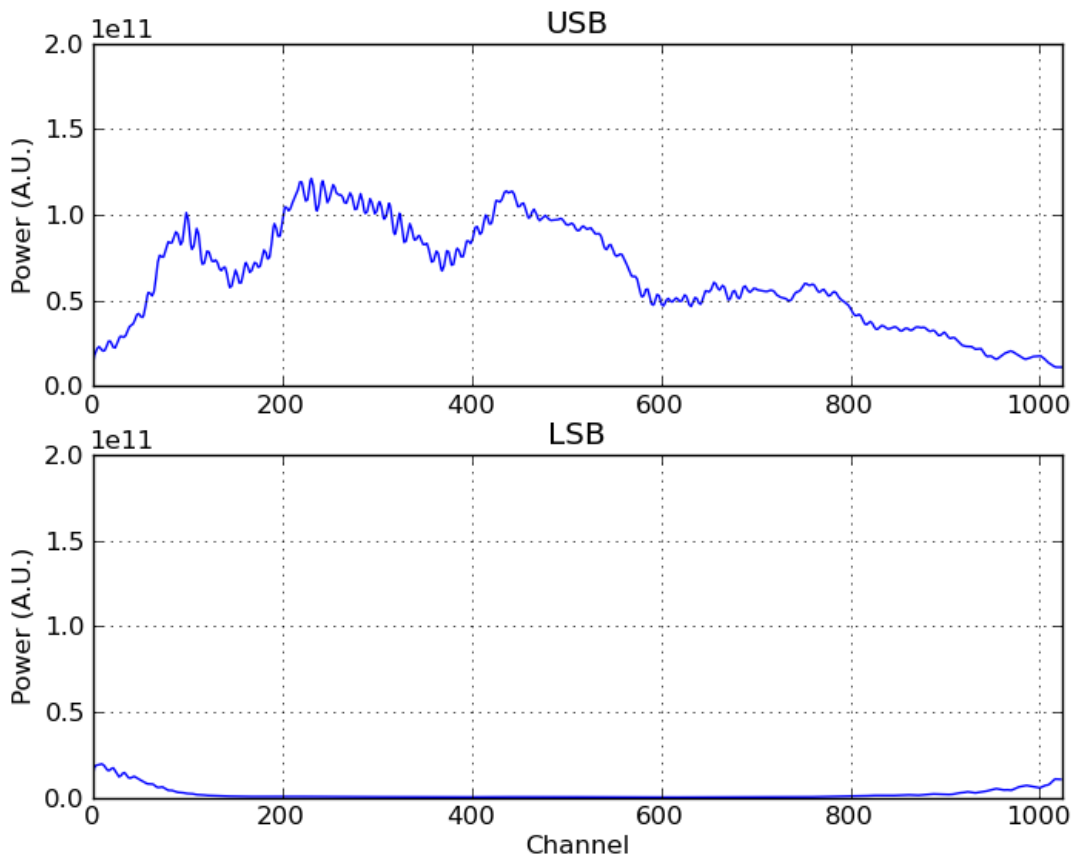

Figure 6. With a band-limited noise in USB provided to the analog down converter, using the nominal coefficients for the digital hybrid, the noise appeared in the USB output after power and delay adjustments.

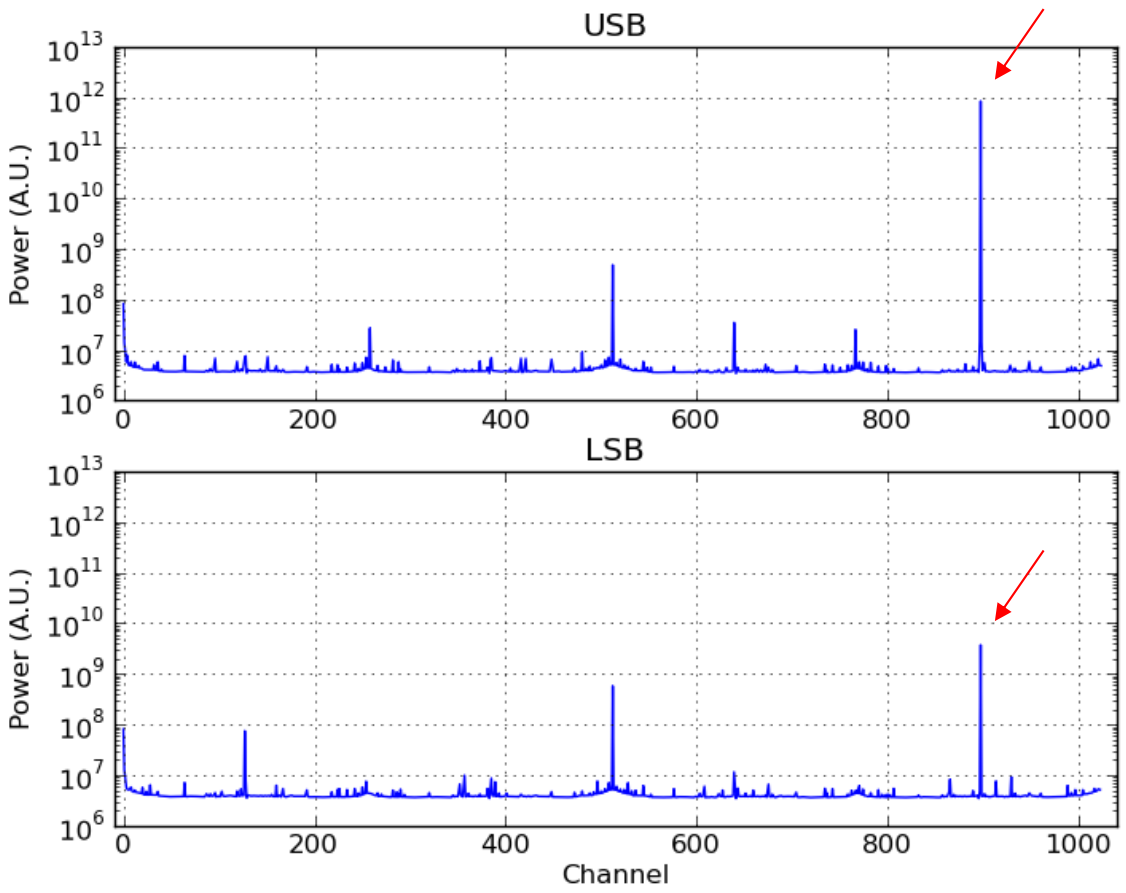

Figure 7. Un-calibrated spectra. For a test tone in the USB, an image signal can be seen in the LSB. Other spurious signals generated in the ADC can also been seen. The sideband rejection ratio here is above 20 dB.

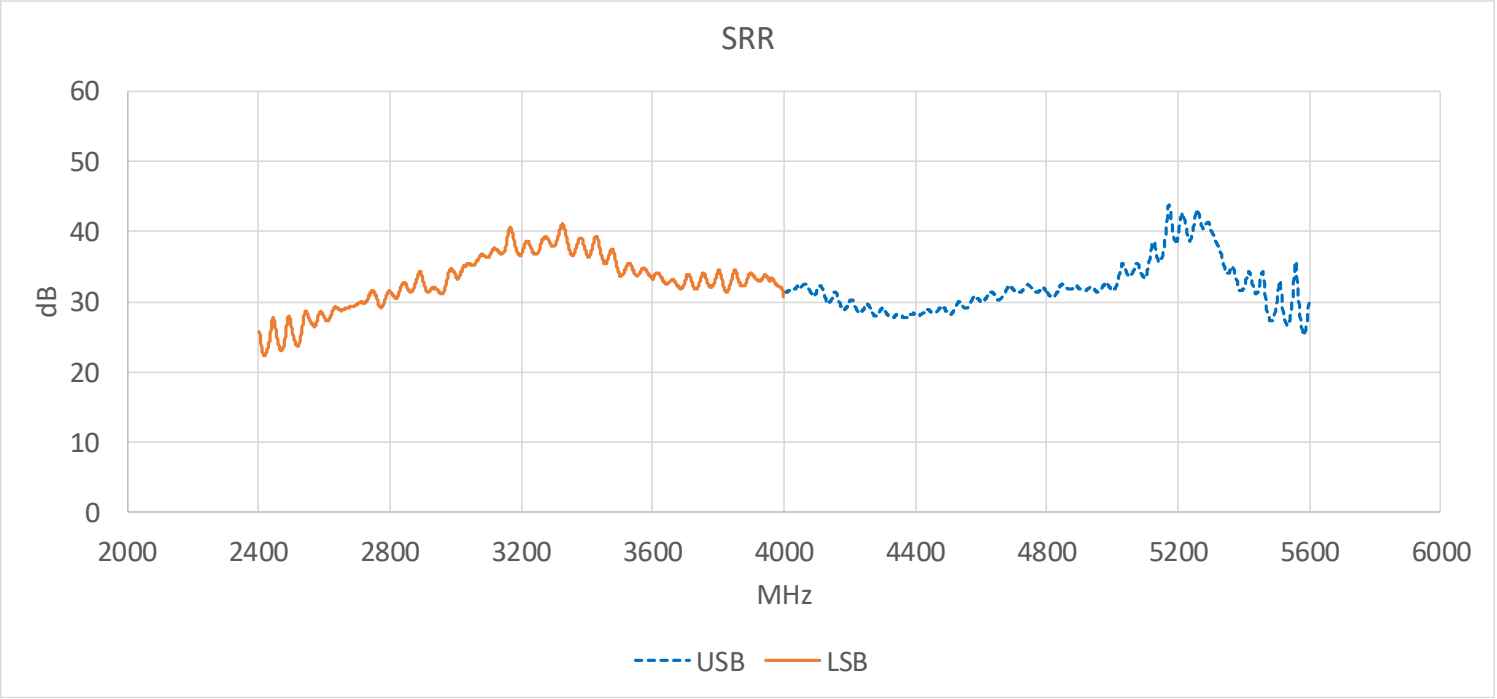

Figure 8. Un-calibrated SRRs across two sidebands as the LO of the analog down converter module is set to 4 GHz.

To calibrate the amplitude and phase imbalances, a test tone was used and correlation was performed after the PFB/FFT to extract the amplitude ratio $X$ and the phase difference $\varphi$ between channels I and Q, as shown in Fig. 9. When a USB signal is present, the phase of the cross-correlation is $\varphi_{USB}$, close to $-\frac{\pi}{2}$, and the power ratio between the two auto-correlations is $X_{USB}^2$. Similarly, for a tone in the LSB, the phase of the cross-correlation is $\varphi_{LSB}$, close to $\frac{\pi}{2}$, and the power ratio is $X_{LSB}^2$. To cancel the image signals, the ratios of the coefficients can be obtained as [7]

$$\frac{C_1}{C_2} = \frac{1}{X_{LSB}} e^{-j(\varphi_{LSB}-\pi)} \tag{1}$$

and

$$\frac{C_3}{C_4} = \frac{1}{X_{USB}} e^{-j(\varphi_{USB}-\pi)}. \tag{2}$$

The complex calibration coefficients for $C_2$ and $C_3$ as $C_1$ and $C_4$ are set to $1 + 0\,j$ are shown in Fig. 10. After the coefficients were applied, the image signal was further reduced, as shown in Fig. 11, and the SRRs increased above 30 dB across the two sidebands, as shown in Fig. 12.

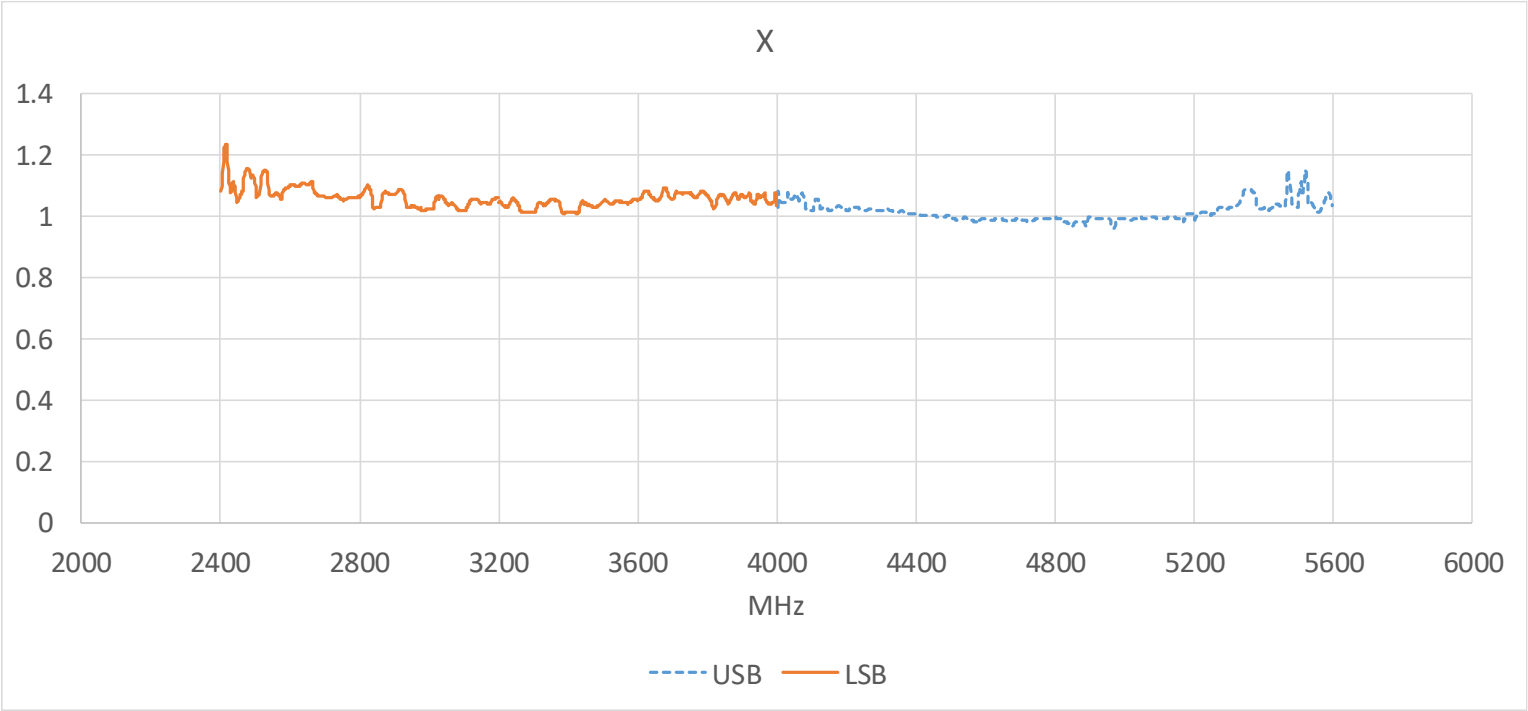

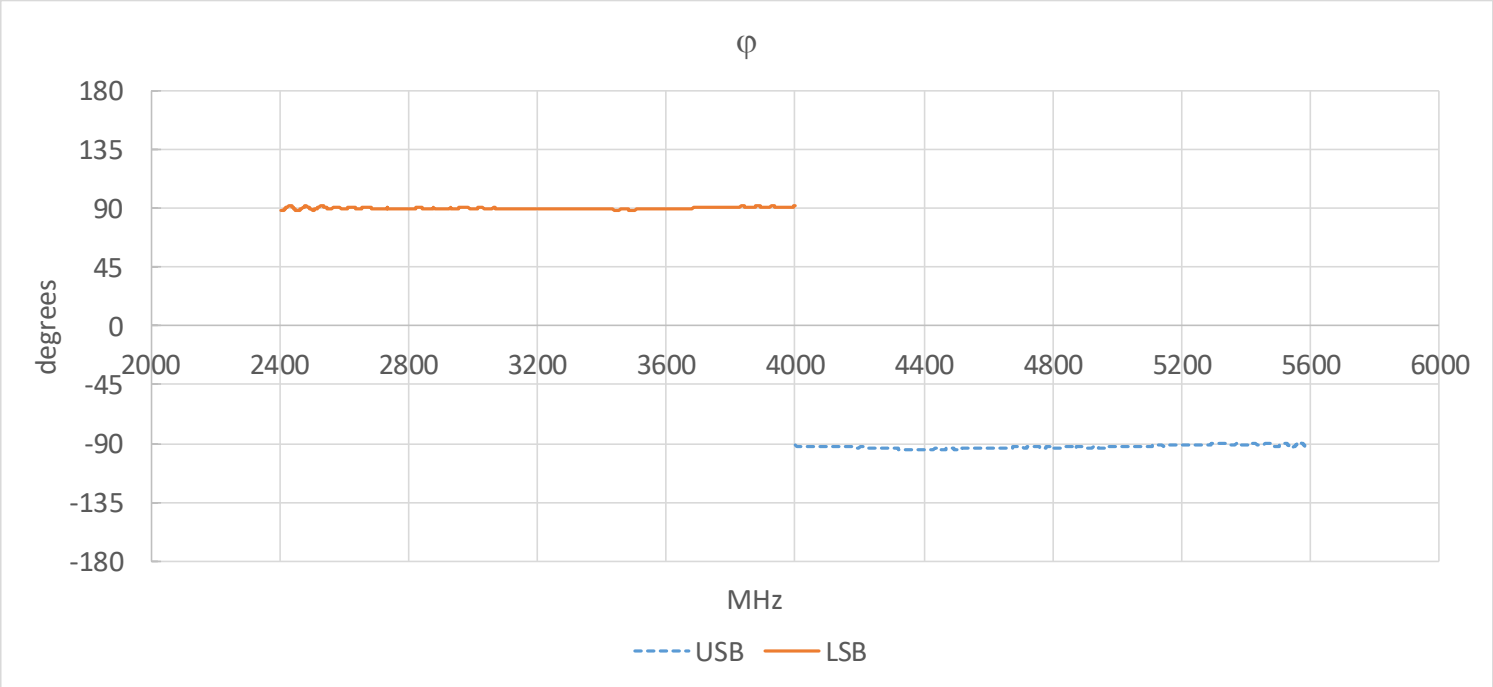

Figure 9. Amplitude ratios and phase differences measured across the two sidebands after amplitude and delay adjustments.

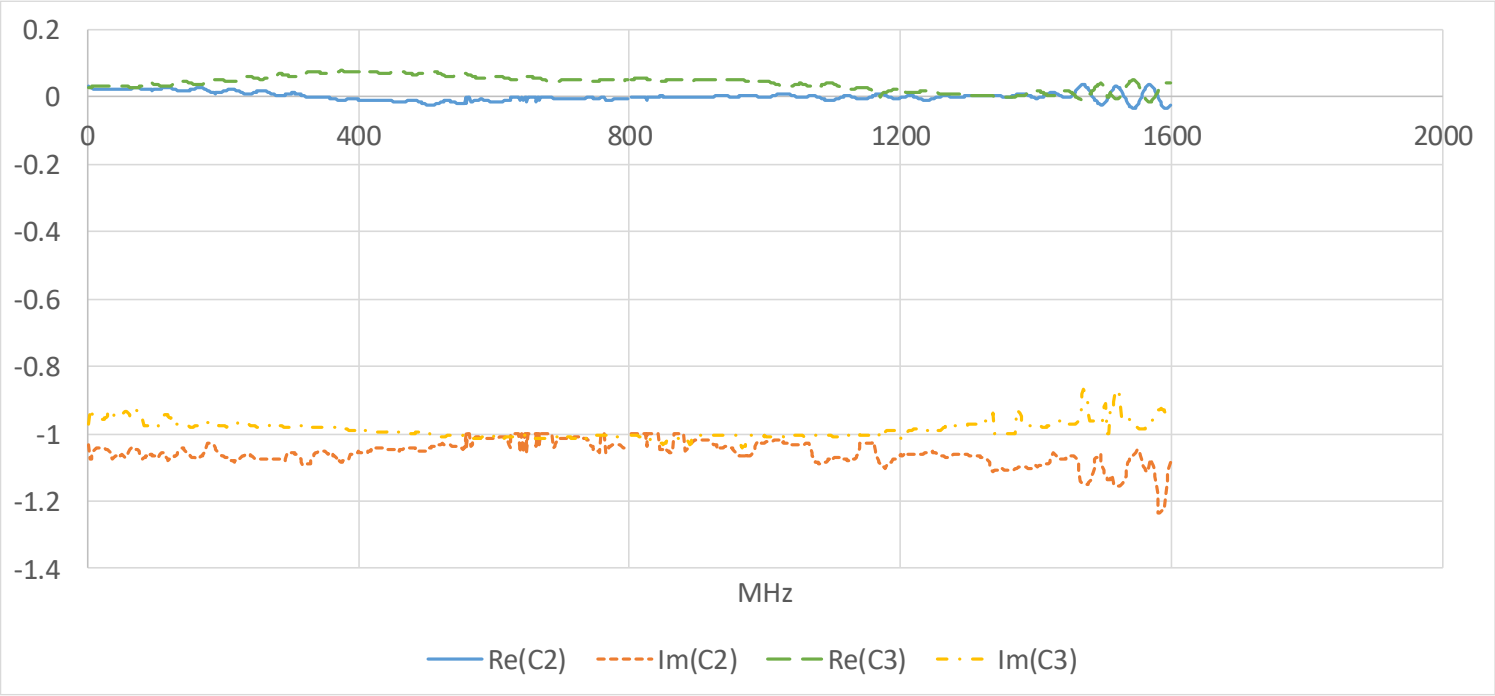

Figure 10. Real and imaginary parts of the complex coefficients $C_2$ and $C_3$, derived from amplitude ratios and phase differences.

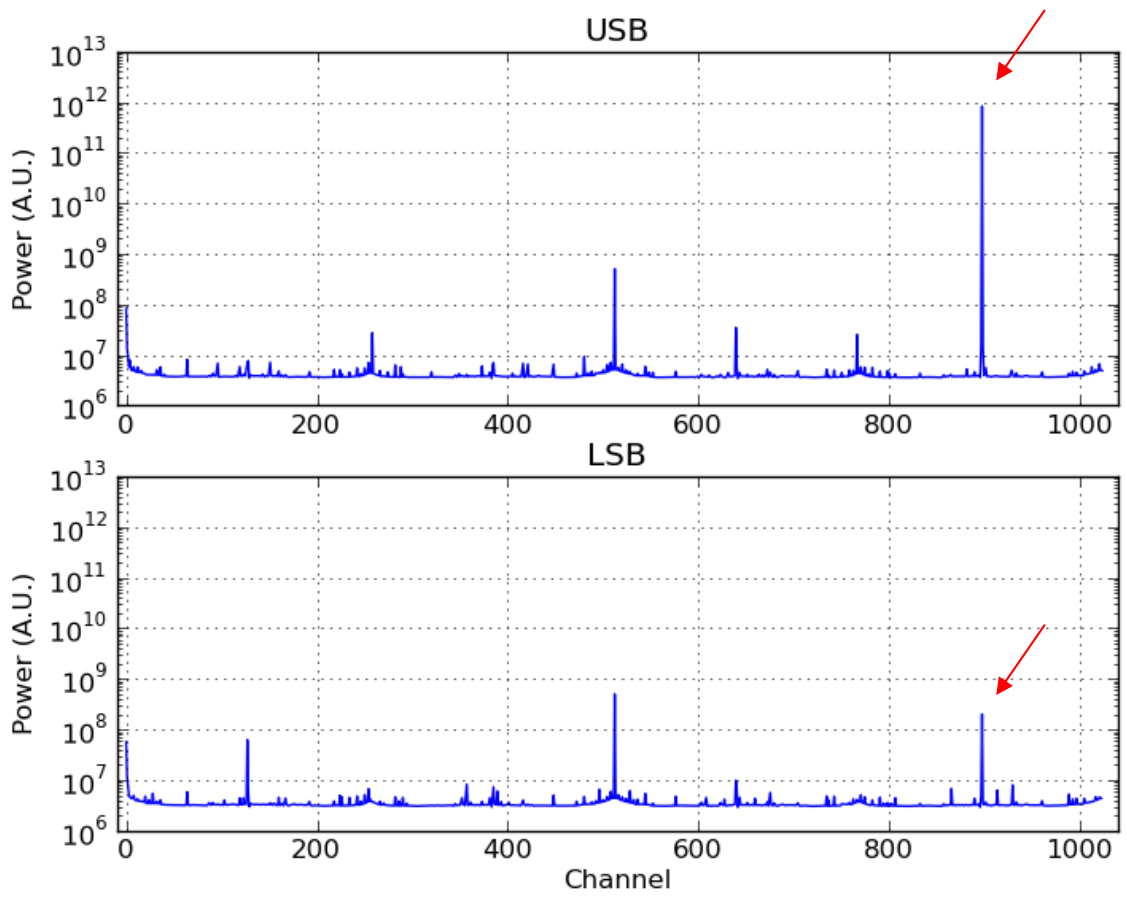

Figure 11. Calibrated spectra with a test tone in the USB. The image signal in the LSB is further suppressed.

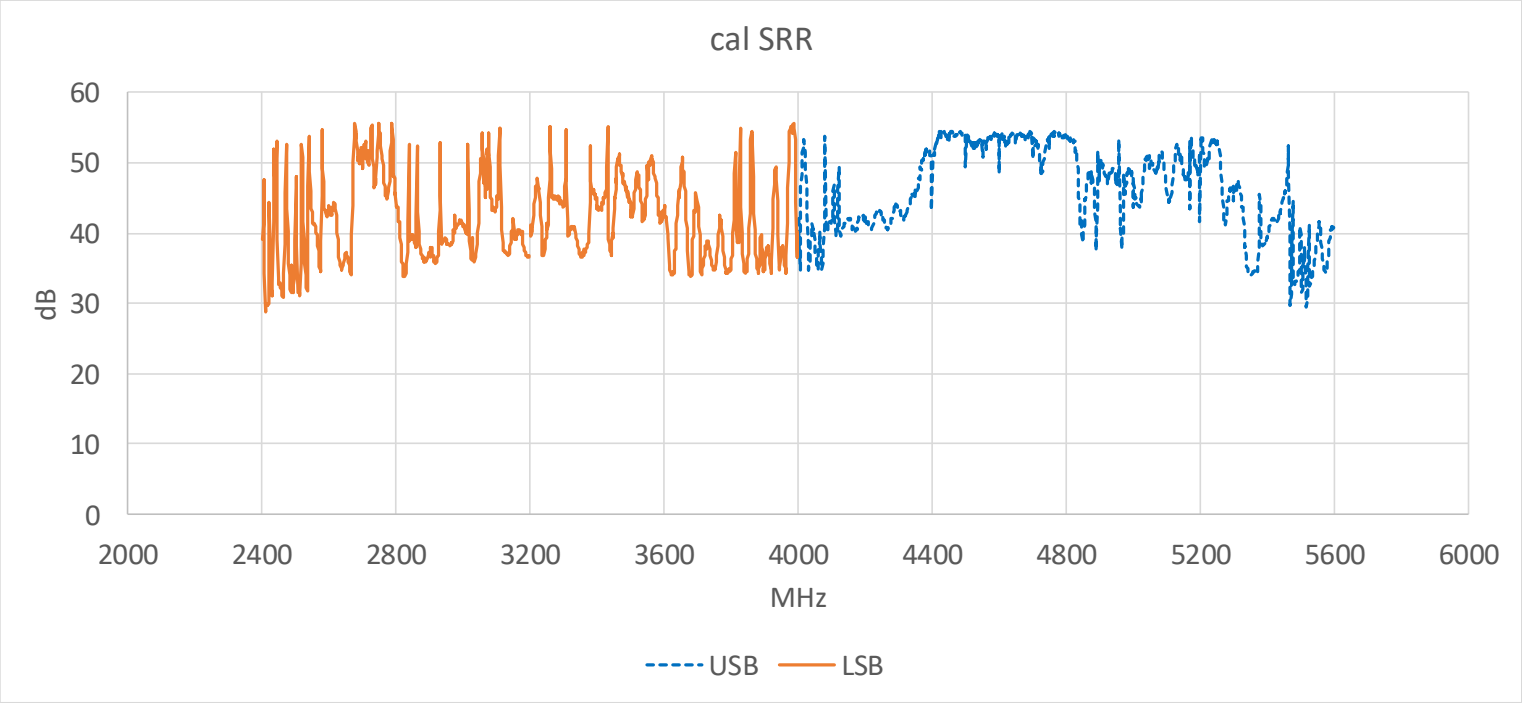

Figure 12. Calibrated SRRs. The ratios are above 30 dB across two sidebands.

## 4. DISCUSSION

The laboratory tests demonstrate that with calibration, the digital sideband-separating down-conversion technique can process a 2 × 1.6 GHz bandwidth with SRRs greater than 30 dB. The bandwidth is currently limited by timing within FPGAs. Another demux by 2 in the ADC interface block will be implemented to resolve this issue. The SRRs are affected by standing waves or multiple reflections, especially between the analog down converter outputs and the ADCs. With the current test setup, the standing wave effect showed a period of approximately 50 MHz, corresponding to coaxial cables with a length of approximately 2.1 m. Future work will be conducted to mitigate the standing wave effect to improve the SRRs.

Large variations in the SRRs and calibration coefficients near the cutoff of the anti-aliasing filters result from rapid changes in the amplitude response and group delay of the filters. Other types of filters (such as maximally flat or linear phase) might be considered for reducing the calibration density near the cutoff and improving the system stability [7].